%% file: IEEE_BigData_2019_paper.tex
\documentclass[conference]{IEEEtran}
\IEEEoverridecommandlockouts
\usepackage{cite}
\usepackage{amsmath,amssymb,amsfonts}
\usepackage{mathtools}
\usepackage{algorithmic}
\usepackage{textcomp}
\usepackage{xcolor}

\usepackage{booktabs} 
\usepackage{etoc} 
\usepackage{multirow}
\usepackage{makecell}
\usepackage{enumitem}

\usepackage{graphicx} 
\usepackage[export]{adjustbox}
\usepackage{subfig}
\usepackage{epstopdf}
\DeclareGraphicsExtensions{.pdf,.png,.eps,.jpg,.tif,.tiff,.ps}
\graphicspath{{img/temp_images//}{img//}}

\usepackage[hyphens]{url}
\usepackage{hyperref} 
\hypersetup{breaklinks=true} 
\usepackage[nameinlink,capitalize]{cleveref}

\usepackage{xcolor}
\usepackage{soul}
\definecolor{navy}{rgb}{0.0, 0.0, 0.0}
\definecolor{ruby}{rgb}{0.0, 0.0, 0.0}
\definecolor{ao}{rgb}{0.0, 0.0, 0.0}

\usepackage[colorinlistoftodos,prependcaption,textsize=tiny]{todonotes}

\def\BibTeX{{\rm B\kern-.05em{\sc i\kern-.025em b}\kern-.08em
    T\kern-.1667em\lower.7ex\hbox{E}\kern-.125emX}}
\begin{document}
\etocdepthtag.toc{mtchapter}


%
\newcommand{\titlename}{Adaptively selecting occupations to detect skill shortages from online job ads}
\title{\titlename}

\makeatletter
\newcommand{\linebreakand}{%
  \end{@IEEEauthorhalign}
  \hfill\mbox{}\par
  \mbox{}\hfill\begin{@IEEEauthorhalign}
}
\makeatother


\author{\IEEEauthorblockN{
Nik Dawson\IEEEauthorrefmark{1}\IEEEauthorrefmark{2},
Marian-Andrei Rizoiu\IEEEauthorrefmark{3}\IEEEauthorrefmark{4},
Benjamin Johnston\IEEEauthorrefmark{2},
Mary-Anne Williams\IEEEauthorrefmark{2}}
\IEEEauthorblockA{
\IEEEauthorrefmark{2}\textit{Centre of Artificial Intelligence, University of Technology Sydney},\\ 
\IEEEauthorrefmark{3}\textit{Faculty of Engineering \& IT, University of Technology Sydney}, \IEEEauthorrefmark{4}\textit{CSIRO's Data61},
Sydney, Australia\\
Email: nikolas.j.dawson@student.uts.edu.au
\IEEEauthorrefmark{1}Corresponding author.
}}

\maketitle

\begin{abstract}
Labour demand and skill shortages have historically been difficult to assess given the high costs of conducting representative surveys and the inherent delays of these indicators.
This is particularly consequential for fast developing skills and occupations, such as those relating to Data Science and Analytics (DSA).
This paper develops a data-driven solution to detecting skill shortages from online job advertisements (ads) data.
We first propose a method to generate sets of highly similar skills based on a set of seed skills from job ads. 
This provides researchers with a novel method to adaptively select occupations based on granular skills data.
Next, we apply this adaptive skills similarity technique to a dataset of over 6.7 million Australian job ads in order to identify occupations with the highest proportions of DSA skills. This uncovers 306,577 DSA job ads across 23 occupational classes from 2012-2019. 
Finally, we propose five variables for detecting skill shortages from online job ads: (1) posting frequency; (2) salary levels; (3) education requirements; (4) experience demands; and (5) job ad posting predictability. This contributes further evidence to the goal of detecting skills shortages in real-time. In conducting this analysis, we also find strong evidence of skills shortages in Australia for highly technical DSA skills and occupations. These results provide insights to Data Science researchers, educators, and policy-makers from other advanced economies about the types of skills that should be cultivated to meet growing DSA labour demands in the future.
\end{abstract}


\section{Introduction}
\label{sec:intro}

The Internet has become the primary channel for disseminating information in many areas of society. This is the case for job advertisements (ads), where approximately 60\% of Australian job ads are posted online~\cite{Department_of_Employment_Skills_Small_and_Family_Business_undated-qv}. At aggregate levels, online job ads can provide valuable indicators of relative labour demands. Rather than relying solely on lagging indicators from labour market surveys, online job ads data can reveal shifting labour demands as they occur. 
This can provide policy-makers, researchers, and businesses with additional data points to assess the health and dynamics of labour markets.

Real-time labour demand data is essential for Data Science and Analytics (DSA) occupations because of how rapidly DSA skills are evolving and diffusing into other occupational classes. In this research, DSA skills refer to the use of scientific methods, processes, algorithms, and systems to extract knowledge and insights from structured and unstructured data, which can be used to make data-driven decisions and actions~\cite{Dhar2013-rd}. DSA skills are multi-disciplinary, adopting methods from fields such as statistics, mathematics, and computer science. A distinction can also be made between skills, knowledge, abilities, and occupations. `Skills' are the proficiencies developed through training and/or experience~\cite{Oecd2019-cl}; `knowledge' is the theoretical and/or practical understanding of an area; `ability' is the competency to achieve a task~\cite{Gardiner2018-dt}; and `occupations' are the amalgamation of skills, knowledge, and abilities that are used by an individual to perform a set of tasks that are required by their vocation. For simplicity, throughout this paper the term `skill' will include `knowledge' and `ability'.

There are several challenges when analysing the labour demands of occupations and assessing the extent of skills shortages. 
The first challenge concerns accurately identifying occupations based on their evolving skill demands. 
Occupations are organised into standardised hierarchical classifications, which vary across national jurisdictions. 
Most often, these are static, rarely updated classifications, which fail to capture the changing skill demands, or to detect the creation of new occupations. 
For instance, `Data Scientists', `Data Engineers' and `Data Analysts' do not exist in the Australian and New Zealand Standard Classification of Occupations (ANZSCO); 
rather, they are all grouped as `ICT Business Analysts'. 
Furthermore, even when occupations are analysed based on their skill frequencies~\cite{Gardiner2018-dt}, biases emerge from the difference in their relative frequency.
For example, `Communication Skills' occur in around one-quarter of all job ads used in this work. 
However, just because some skills are common does not mean that they are more or less important than other skills that are also required in an individual job. 
This leads to two related questions: 
(1) \textbf{how to adaptively identify relevant skills from labour market data while minimising biases that emerge from ad hoc aggregations?} And 
(2) \textbf{how to identify relevant occupations based on this generated set of skills?}

The second challenge is detecting evidence of skills shortages from (near) real-time data. 
Skill shortages are mostly measured via labour market surveys~\cite{Nedelkoska2015-jm}. 
This involves surveying employers about their abilities to access workers who possess the skills their firms demand. 
A major shortcoming of this approach is that surveys are difficult to scale, and that they are rarely conducted on statistically valid samples~\cite{Department_of_Employment_undated-am}. 
Another significant issue is that labour market surveys are lagging indicators, i.e. the publication of results can be many months after the data was collected. 
Lastly, due to scaling limitations, prominent labour market surveys on skills shortages (or mismatches) fail to measure all standardised occupations~\cite{Oecd2019-cl}. 
Therefore, the questions are \textbf{can we detect evidence of skill shortages from real-time labour market data? If so, what are the key variables for assessing skills shortages from such data?}

This paper addresses the above challenges using a large dataset of over 6.7 million Australian online job ads spanning between 2012-01-01 and 2019-02-28, which has been generously provided by Burning Glass Technologies\footnote{BGT is a leading vendor of online job ads data. \texttt{https://www.burning-glass.com/}} (BGT). 
The data has been collected via web scraping and systematically processed into structured formats. The dataset consists of detailed information on individual job ads, such as location, salary, employer, educational requirements, experience demands, and more. The skill requirements have also been extracted (totalling $>11,000$ unique skills) and each job ad is classified into its relevant occupational and industry classes.

To address the first challenge, we first adapt an established similarity measure originating from Trade Economics~\cite{Hidalgo2007-qk} to measure the pairwise similarity between unique skills in job ads.
Next, we develop a novel data-driven method to generate sets 
of skills highly similar to a set of seed skills.
Finally, we uncover the relevant occupations for which at least $15\%$ of all skills required in their associated ads are from the target set of skills.
We apply this method to uncover the set of DSA skills and DSA occupations, starting from a seed set of common DSA skills.


We address the second challenge by identifying five key variables from online job ads data which are critical for detecting skill shortages in real-time:
(1) job ad posting frequency; (2) median salary levels; (3) educational requirements; (4) experience demands; and (5) job posting predictability. 
We then analyse the DSA occupations according to each of these five variables and find compelling evidence for how these features are predictive of skill shortages.

\textbf{The main contributions of this work include:}
\begin{itemize}
    \item We develop a \textbf{data-driven methodology to construct skills sets} for specific occupational areas, and to select occupations based on granular skills-level data;
    \item We identify \textbf{five key variables for detecting skill shortages from online job ads data};
    \item We apply the aforementioned methods to a unique dataset of online job ads to \textbf{analyse the changing labour demands of DSA skills and occupations} in the advanced economy of Australia. We also \textbf{construct and share the list of top DSA skills} generated from this dataset.
\end{itemize}

\section{Related Work \& Limitations} \label{related-work}

\textbf{Job ads data as a proxy for labour demand.}
During 2001-2003, Lee~\cite{lee2005analysis} gathered job ads data from the websites of Fortune 500 companies in order to analyse the skill requirements of `Systems Analysts'. 
Lee was able to determine that these positions demanded their candidates to have `all-round' capabilities, beyond just technical skills.
More recently, Gardiner et al.~\cite{Gardiner2018-dt} procured 1,216 job ads with `Big Data' in the job title from the \texttt{indeed.com} API. 
The authors then conducted content analyses to investigate how `Big Data' skills have manifested in labour demand. 
Their research reiterated that employers are demanding technical skills in conjunction with `softer' skills, such as communication and team-work.

\textbf{DSA skill shortages.}
While the capacity to collect, store, and process information may have sharply risen, it is argued that these advances have far outstripped present capacities to analyse and make productive use of such information~\cite{Hey2009-hj}. Claims of DSA skill shortages are being made in labour markets around the world~\cite{Blake2019-ha,LinkedIn_Economic_Graph_Team2018-gu,Manyika2011-vh}, including in Australia~\cite{appendix}.
Most similar to this research, however, are two studies conducted using BGT data to assess DSA labour demands. The first was an industry research collaboration between BGT, IBM, and the Business-Higher Education Forum in the US~\cite{Markow2017-hu}. The research found that in 2017 DSA jobs earned a wage premium of more than US\$8,700 and DSA job postings were projected to grow 15\% by 2020, which is significantly higher than average. In another study commissioned by the The Royal Society UK~\cite{Blake2019-ha}, BGT data were analysed for DSA jobs in the UK. 
The results also showed high levels of demand for DSA skills, particularly `technically rigorous' DSA skills.

\textbf{Limitations of using online job ads data.}
It is argued that job ads data are an incomplete representation of labour demand. Some employers continue to use traditional forms of advertising for vacancies, such as newspaper classifieds, their own hiring platforms, or recruitment agency procurement. Job ads data also over-represent occupations with higher-skill requirements and higher wages, colloquially referred to as `white collar' jobs~\cite{appendix, Carnevale2014-xc}.

\textbf{Occupational classifications.}
There are significant shortcomings job ads data that are classified according to official occupational standards.
Official occupational classifications, like ANZSCO, are often static taxonomies and are rarely updated. We therefore use the BGT occupational classifications because of its adaptive taxonomies that update with changing labour demands. For example, a job ad title of `Senior Data Scientist' is classified as a `Data Scientist' in the BGT occupational classification but is classified as an `ICT Business \& Systems Analyst' by ANZSCO. For more details, please review the online appendix~\cite{appendix}.

\section{Skill similarity and sets of related skills}
\label{sec:skills-similarity}

\textbf{Intuition.}
Skills provide the means for workers to perform labour tasks in order to fulfill their occupational demands. Therefore, the assortment of skills required for a job, and their pairwise interconnections uniquely identify occupations. 
In this section, we propose a methodology to capture the `similarity' between skill-pairs that co-occur in job ads. 
Intuitively, two skills are similar when the two are related and complementary, i.e. the skills-pair supports each other. 
For example, `Python' and `TensorFlow' have a high similarity score because together they enable higher productivity for the worker, and because the difficulty to acquire either skill when one is already possessed by a worker is relatively low.

\textbf{The Revealed Comparative Advantage of a skill.}
We develop a data-driven methodology to measure the pairwise similarity between pairs of skills that co-occur in job ads. 
One difficulty we encounter is that some skills are ubiquitous, occurring across many job ads and occupations. 
We address this issue by adapting the methodology proposed by Alabdulkareem et al.~\cite{Alabdulkareem2018-jl} to maximise the amount of skill-level information obtained from each job ad, while minimising the biases introduced by over-expressed skills in job ads.
We use the \textit{Revealed Comparative Advantage} (RCA) to measure the relevance of a skill $s$ for a particular job ad $j$, computed as:
\begin{equation*}
  RCA(j, s) = \frac{x(j, s) / \mathop{\sum}\limits_{s'\in \mathcal{S}}x(j, s')}
  {\mathop{\sum}\limits_{j'\in J}x(j', s) / \mathop{\sum}\limits_{j'\in \mathcal{J},s'\in \mathcal{S}}x(j', s')}
\end{equation*}
\noindent
where $x(j,s) = 1$ when the skill $s$ is required for job $j$, and $x(j,s) = 0$ otherwise;
$\mathcal{S}$ is the set of all distinct skills, and $\mathcal{J}$ is the set of all job ads in our dataset.
$RCA(j, s) \in \left[ 0, \mathop{\sum}\limits_{j'\in J,s'\in S}x(j', s') \right], \forall j, s$, and the higher $RCA(j, s)$ the higher is the comparative advantage that $s$ is considered to have for $j$.
Visibly, $RCA(j, s)$ decreases when the skill $s$ is more ubiquitous (i.e. when $\mathop{\sum}\limits_{j'\in J}x(j', s) $ increases), or when many other skills are required for the job $j$ (i.e. when $\mathop{\sum}\limits_{s'\in S}x(j, s')$ increases).

$RCA$ provides a method to measure the importance of a skill in a job ad, relative to the total share of demand for that skill in all job ads. It has been applied across a range of disciplines, such as trade economics~\cite{Hidalgo2007-qk}~\cite{Vollrath1991-kr}, identifying key industries in nations~\cite{Shutters2016-fe}, and detecting the labour polarisation of workplace skills~\cite{Alabdulkareem2018-jl}.

\textbf{Measure skill similarity.}
The next step is measuring the complementarity of skill-pairs that co-occur in job ads. 
First we introduce the `effective use of skills'
$e(j, s)$ defined as $e(j, s) = 1 \text{ when } RCA(j,s) > 1$ and $e(j, s) = 0$ otherwise.
Finally, we introduce the skill complementarity (denoted $\theta$) as the minimum of the conditional probabilities of a skills-pair being effectively used within the same job ad. 
Skills $s$ and $s'$ are considered as highly complementary if they tend to commonly co-occur within individual job ads, for whatever reason. Formally:
\begin{equation*}
  \theta(s, s') = \frac{\mathop{\sum}\limits_{j'\in J}e(j,s).e(j,s')}
  {max \left( \mathop{\sum}\limits_{j'\in J}e(j,s), \mathop{\sum}\limits_{j'\in J}e(j,s') \right)}
\end{equation*}
Note that $\theta(s, s') \in [0, 1]$, a larger value indicates that $s$ and $s'$ are more similar, and it reaches the maximum value when $s$ and $s'$ always co-occur (i.e. they never appear separately).

\textbf{Top DSA skills.}
We use the $\theta$ function to create a list of DSA skills.
First, we qualitatively select 5 common DSA skills as seed inputs: \textit{`Artificial Intelligence', `Big Data', `Data Mining', `Data Science'}, and \textit{`Machine Learning'}. 
Next, for each of these 5 DSA skills, we calculate the top 300 skills with the highest similarity scores. 
Finally, we merge the five lists, we calculate the average similarity scores for each unique skill, and rank in descending order. 
This results in a ranked list of 589 skills, which we qualitatively assess and decide keep the top 150 skills. 
While some skills outside of the top 150 could be considered DSA skills, it was at this point that the relevance to DSA skills began to deteriorate and merge into other domains. 
For example, skills such as \textit{`Design Thinking', `Front-end Development'}, and \textit{`Atlassian JIRA'} -- which are technical, but not DSA specific -- were just outside of the top 150 skills. 

The purpose of this top DSA skills list is to capture DSA labour trends rather than represent a complete taxonomy of DSA skills. 
The list of top 150 DSA skills can viewed in the online appendix~\cite{appendix}.

\section{DSA occupations and categories}

\textbf{Compute the skill intensity.}
In this section, we present an adaptative technique to uncover DSA occupations from job data.
First, we compute $\eta$ the
\textit{`DSA skill intensity'} for each standardised BGT occupation, defined as percentage of DSA skills relative to the total skill count for the job ads related to an occupation $o$. 
Formally:
\begin{equation*}
	\eta(o, \mathcal{D}) = \frac{\mathop{\sum}\limits_{j \in \mathcal{O}, s \in \mathcal{D}} x(j, s)}{\mathop{\sum}\limits_{j \in \mathcal{O}, s' \in S} x(j, s')}
\end{equation*}
\noindent
where $\mathcal{D}$ is the set of DSA skills, and $\mathcal{O}$ is the set of job ads associated with the occupation $o$.

\textbf{Select the top DSA occupations.}
We qualitatively assessed the occupational list ordered by $\eta$, and decided to establish a cutoff at $\eta > 15\%$. 
The rationale for this threshold level was that occupations just below this cutoff are questionably considered DSA occupations -- take for example, \textit{`Web Developer'} and \textit{`UI / UX Designer / Developer'}. 
Occupations just above this threshold appeared more consistent with the definition of DSA skills given in \cref{sec:intro}. 
Moreover, the occupations with a DSA skill intensity level just above the $15\%$ threshold represented occupations where the authors considered DSA skills to likely become more prevalent. 
For example, the demands for DSA skills are expected to increase for `Economists' due to the growing amounts of economic data that are being made available~\cite{Einav2014-hv}. 
Therefore, this list represents occupations where DSA skills are already important, or have reached a minimum threshold of DSA skill intensity and where DSA skills are likely to become more important for the occupation.

\begin{table}[btp]
	\small
	\caption{
		Selected DSA Occupations and their job ad counts.
	}
    \label{tab:table1}
    \centering
    \begin{tabular}{lp{4.2cm}r}
      \toprule
      \multicolumn{1}{c}{\textbf{DSA Category}} & 
      \multicolumn{1}{c}{\textbf{DSA Occupation}} &
      \multicolumn{1}{r}{\textbf{\#Ads}}\\
      \midrule
      \multirow{10}{2cm}{Data Scientists and Advanced Analysts} 
      & Biostatistician & 270\\
      & Computer Scientist & 38\\
      & Data Engineer & 71\\
      & Data Scientist & 2,388\\
      & Economist & 2,127\\
      & Financial Quantitative Analyst & 947\\
      & Mathematician & 105\\
      & Physicist & 423\\
      & Robotics Engineer & 18\\
      & Statistician & 2,535\\
      \midrule
      \multirow{3}{2cm}{Data Analyst} 
      & Business Intelligence Architect / Developer & 3,166\\
      & Data / Data Mining Analyst & 34,520\\
      \midrule
      \multirow{7}{2cm}{Data Systems Developers}
      & Computer Programmer & 16,311\\
      & Computer Systems Engineer / Architect & 73,437\\
      & Data Warehousing Specialist & 964\\
      & Database Administrator & 17,937\\
      & Database Architect & 7,489\\
      & Mobile Applications Developer & 4,357\\
      & Software Developer / Engineer & 113,247\\
      \midrule
      \multirow{4}{2cm}{Functional Analysts}
      & Business Intelligence Analyst & 23,547\\
      & Fraud Examiner / Analyst & 653\\
      & Security / Defense Intelligence Analyst & 482\\
      & Test Technician & 1,592\\
      \midrule
      \textbf{TOTALS}
      & \textbf{23 DSA Occupations} & \textbf{306,577}\\
      \bottomrule
    \end{tabular}
\end{table}

\cref{tab:table1} shows the 23 occupational classes that satisfy these DSA threshold requirements. 
Occupations are categorised to compare labour dynamics within the DSA occupational set. 
The occupational categories are adapted from previous BGT research completed in the US~\cite{Markow2017-hu} and UK~\cite{Blake2019-ha}. 
Here, BGT grouped DSA occupations into categories based on skill similarities and sorted categories according to `analytical rigour' of their skill sets~\cite{Blake2019-ha}.
We have applied their categorical framework here because (1) we are using the equivalent BGT dataset for the Australian labour market and (2) many of the DSA occupations uncovered in this research are also present and categorised in their studies.
\cref{fig:cat-definitions} illustrates the categorical framework, giving a brief definition of each category and places them on a comparative scale of `analytical rigour'. 

\begin{figure}[tbp]
    \centering
    \includegraphics[width=8cm]{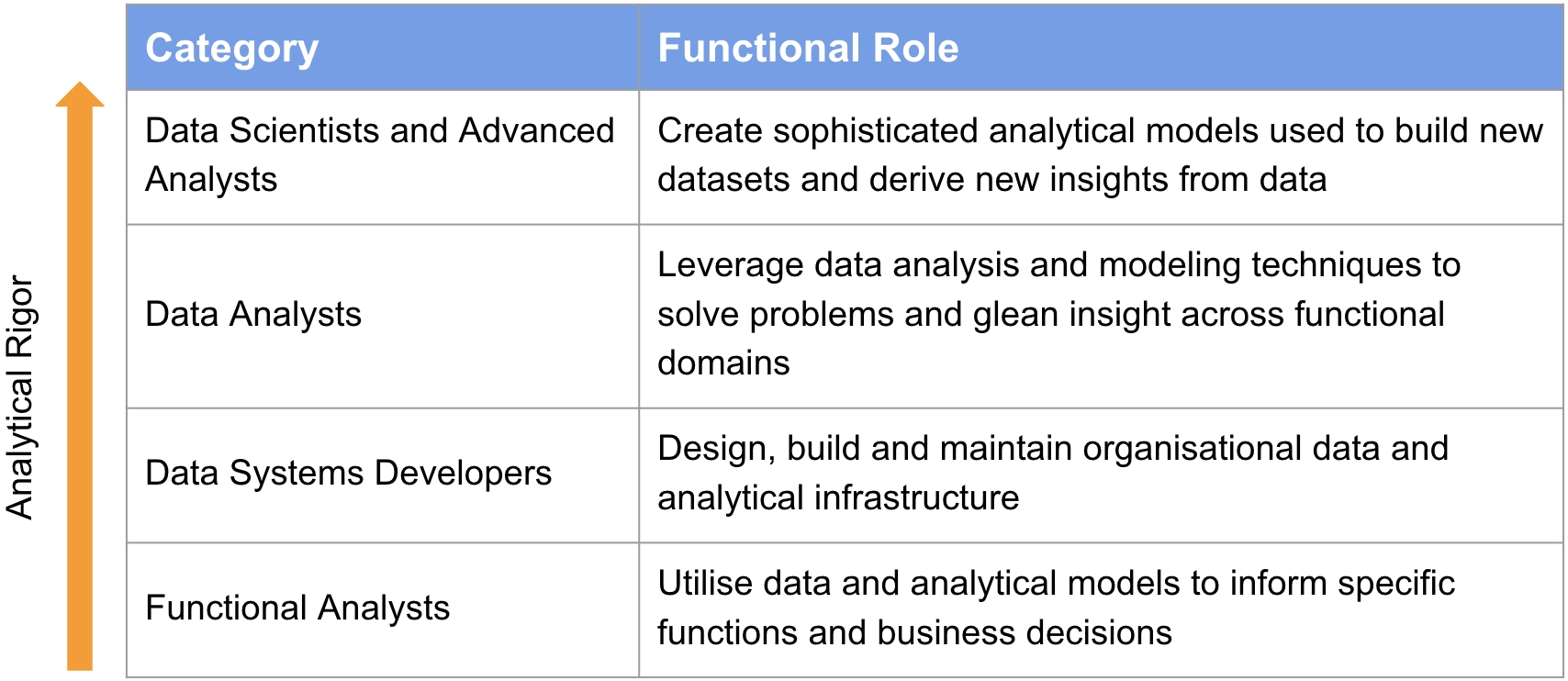}
    \caption{Defining DSA Categories}
    \label{fig:cat-definitions}
\end{figure}

\section{Detecting Skill Shortages from Job Ads}
In this section, we propose five labour demand variables for detecting skill shortages from job ads data. These include: (1) job ad posting frequency growth; (2) median salary levels; (3) educational requirements; (4) experience demands; and (5) job posting predictability. We argue that these variables taken together provide explanatory insight for identifying skill shortages of occupations. 

\subsection{Variables for detecting skill shortages}

This research has found evidence of DSA skill shortages for the `Data Scientists and Advanced Analysts' (`Data Scientists', henceforth) and `Data Analysts' categories. 
A combination of factors have led to these conclusions. 

\textbf{Job ads posting frequency.} 
Both categories have experienced high relative growth in terms of posting frequencies (shown in \cref{subfig:posting-freq}). 
High posting frequency growth can be indicative of increasing employer demands for workers that possess specific occupational skills~\cite{Shinsaku_Nomura_Saori_Imaizumi_Ana_Carolina_Areias_Futoshi_Yamauchi2017-iy}. 
Both `Data Scientists' and `Data Analysts' have averaged higher than average year-on-year growth rates ($28\%$ and $13\%$, respectively) than the other DSA categories and the market average ($10\%$) (see \cref{subfig:posting-growth}).

\textbf{Salaries.} 
`Data Scientists' and `Data Analysts' command high, and growing, wage premiums (\cref{subfig:median-salary}). 
High and growing wages indicate that employers are willing to pay a premium to attract workers with specific skills~\cite{Cappelli2015-wy}. That is, when labour supply is constrained and labour demand increases, then wages should increase, as is the case for `Data Scientists' and `Data Analysts'.

\textbf{Education levels.} 
High relative educational requirements can constrain the supply of skilled labour by creating barriers to entry ~\cite{Cappelli2015-wy}. 
In \cref{subfig:education-level}, this is especially evident for `Data Scientists', where the years of education required by employers is significantly higher than average and other categories.

\textbf{Experience demands.} The minimum years of experience demanded by employers can vary according to the accessibility of skilled labour. If employers have difficulty hiring the labour they demand, then they may reduce their experience-level requirements as part of their recruitment efforts~\cite{Helpman1999-fy}. 
As \cref{subfig:experience} shows, this is again the case for `Data Scientists' and to a lesser extent `Data Analysts', where experience levels have remained relatively low. For `Data Scientists', the minimum experience requirements have decreased by almost one year since 2012 and sit just above the market average. For `Data Analysts', the average years of minimum experience have been below the market average since after 2016. 

\textbf{Job ad posting predictability.} Lastly, we assert that the predictability of job ad posting frequency should be considered as an explanatory variable for detecting skill shortages. We have observed the difficulties of predicting occupations (and skills) that have high-growth in terms of job ad postings. 
As seen in \cref{subfig:post-predictability}, the forecast predictions for `Data Scientists' job ads perform relatively poorly compared to the lower growth categories. 
We contend that this is due to the rapidly changing labour dynamics of `Data Scientists' and that this lack of predictability tends to highlight the patterns of high-growth occupations, reflecting another measure of rising labour demands.
In the next section (\cref{subsec:job-ad-prediction}) we detail how we quantify the predictability variable.

Taken collectively, these factors form a strong case that the Australian labour market has been experiencing a shortage of `Data Scientists' and `Data Analysts'. These variables form a framework of features to detect skill shortages from job ads.

%

\begin{figure*}
	\centering
	\newcommand\myheightA{0.132} 
	\newcommand\myheightB{0.121} 

	\subfloat[]{
		\includegraphics[height = \myheightA\textheight]{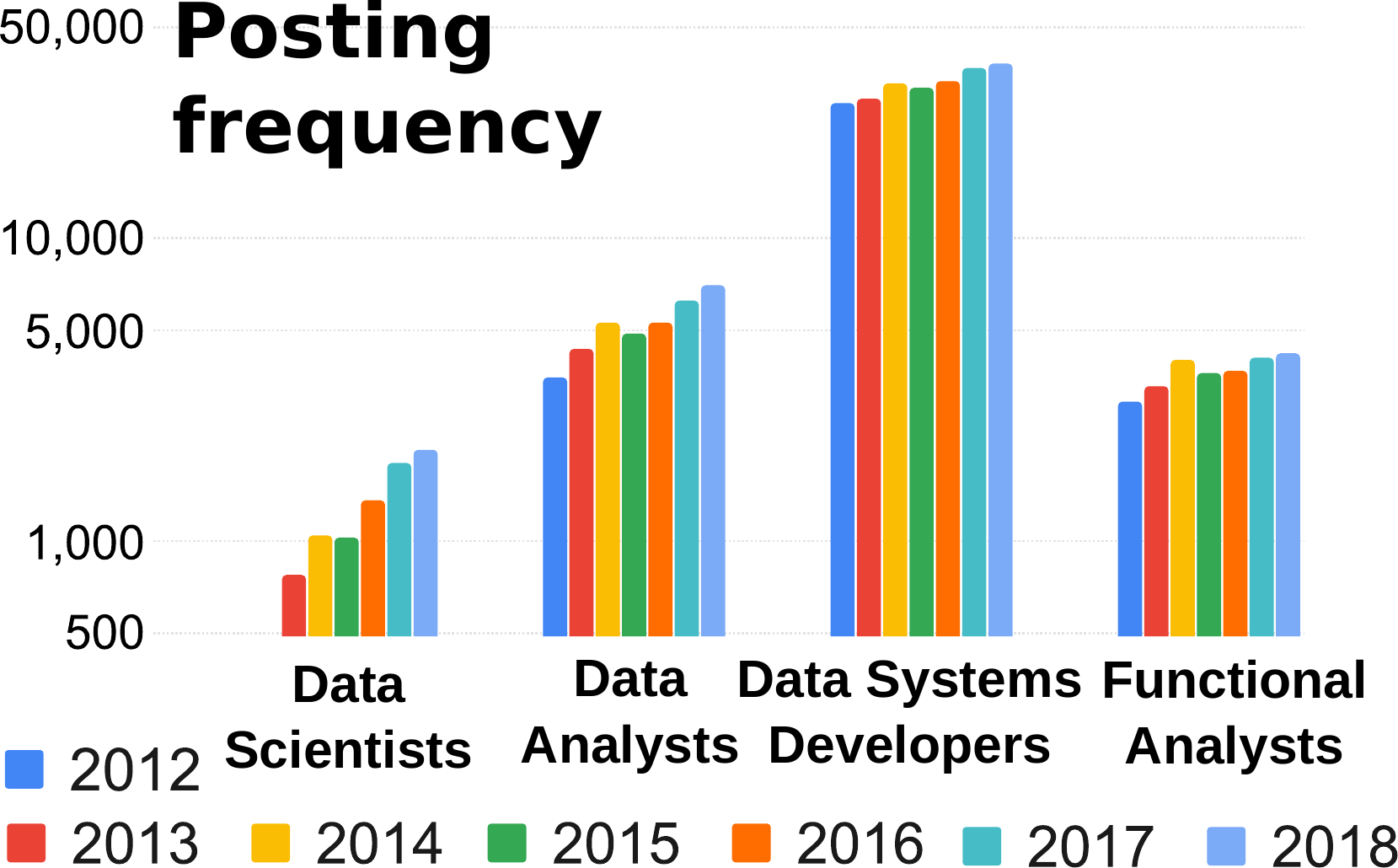}
		\label{subfig:posting-freq}
	}
	\subfloat[]{
		\includegraphics[height = \myheightA\textheight]{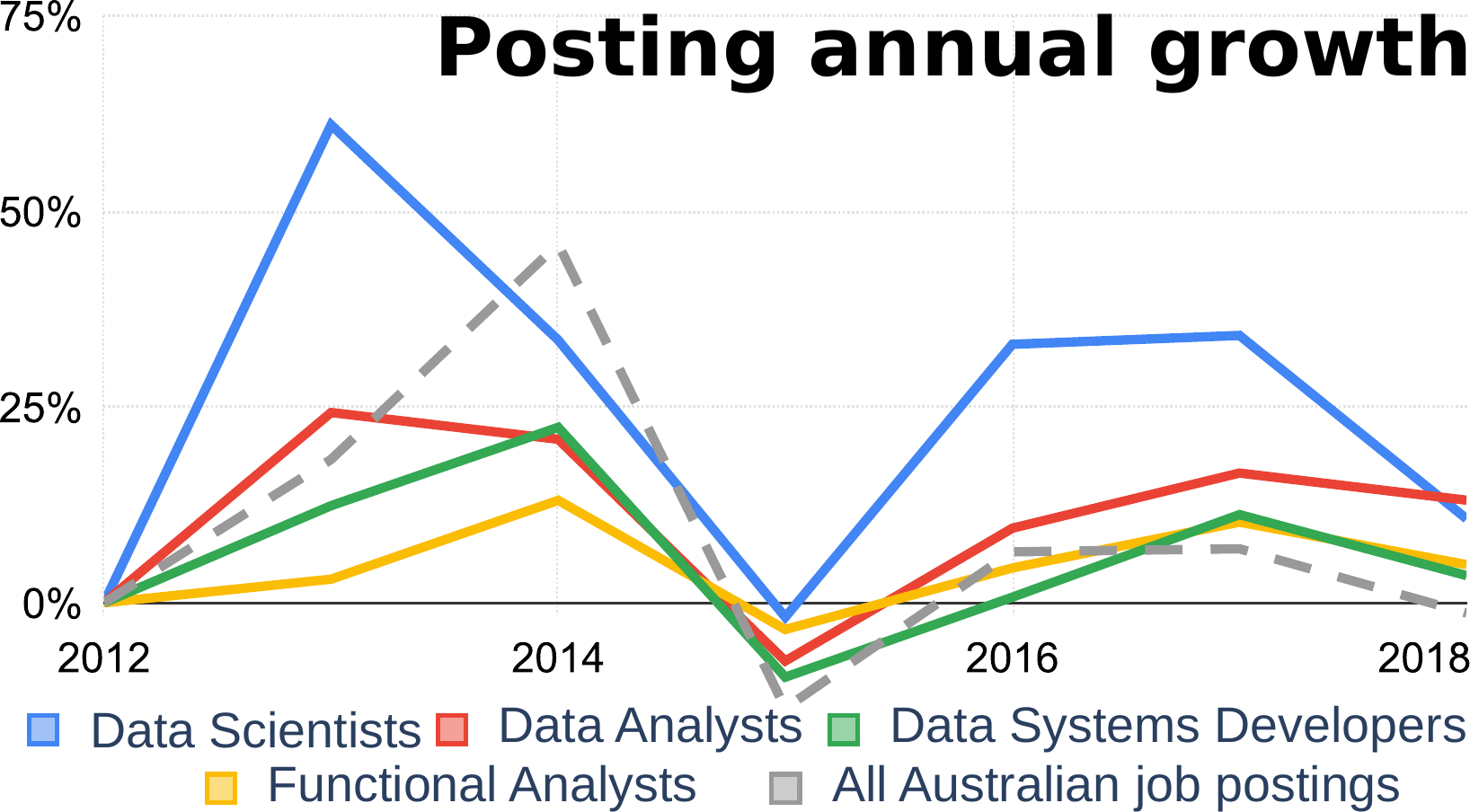}
		\label{subfig:posting-growth}
	}
	\subfloat[]{
		\includegraphics[height = \myheightA\textheight]{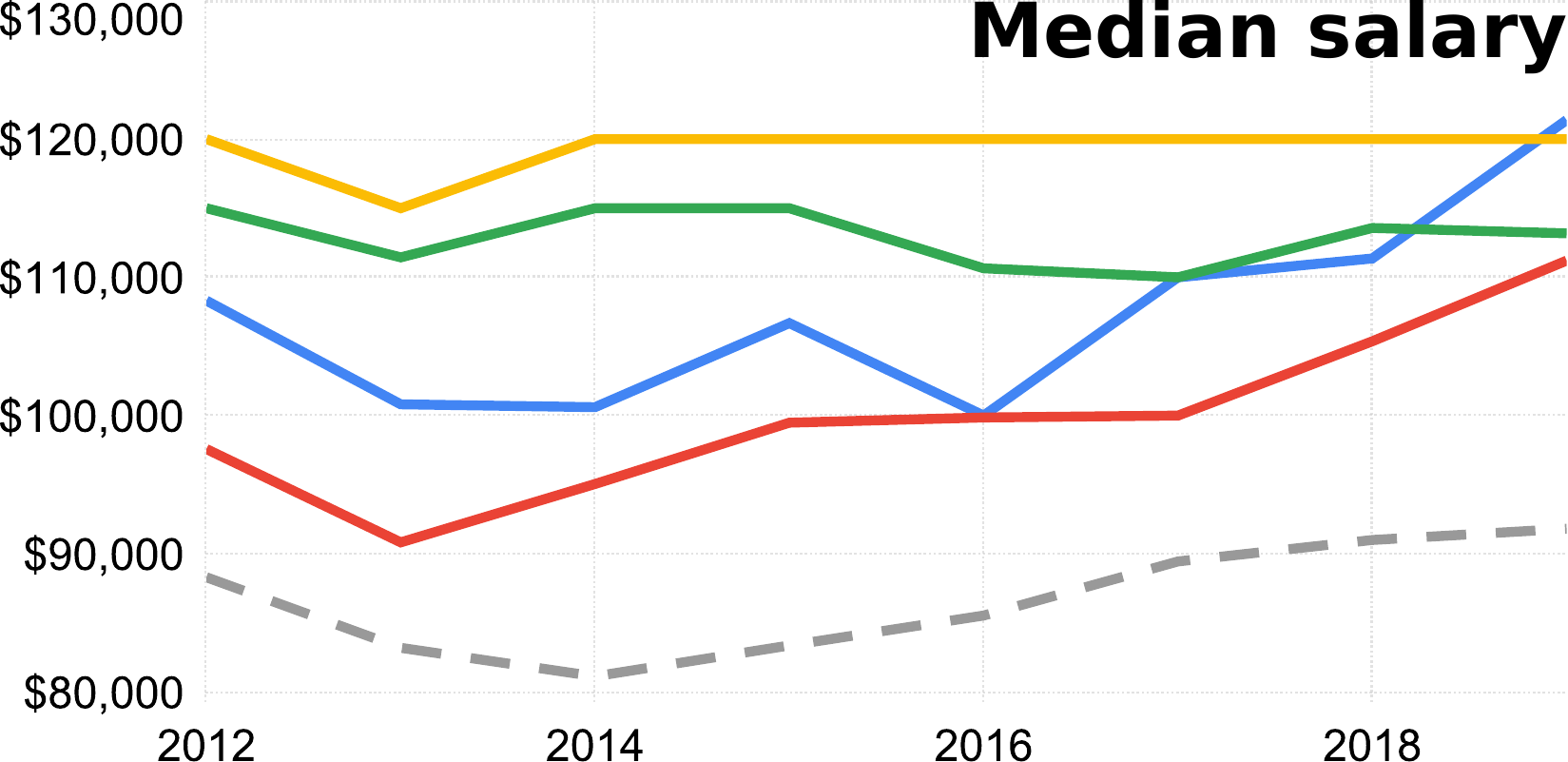}
		\label{subfig:median-salary}
	}\\
	\subfloat[]{
		\includegraphics[height = \myheightB\textheight]{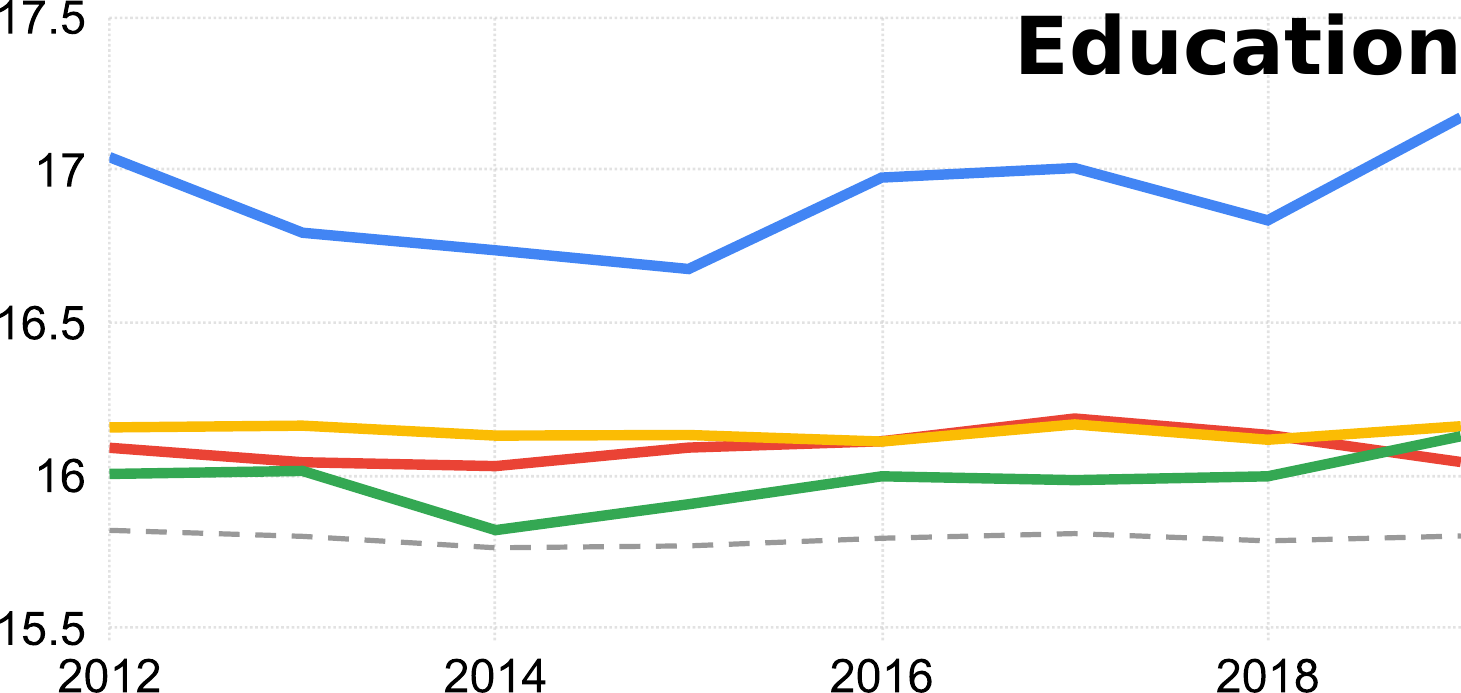}
		\label{subfig:education-level}
	}
	\subfloat[]{
		\includegraphics[height = \myheightB\textheight]{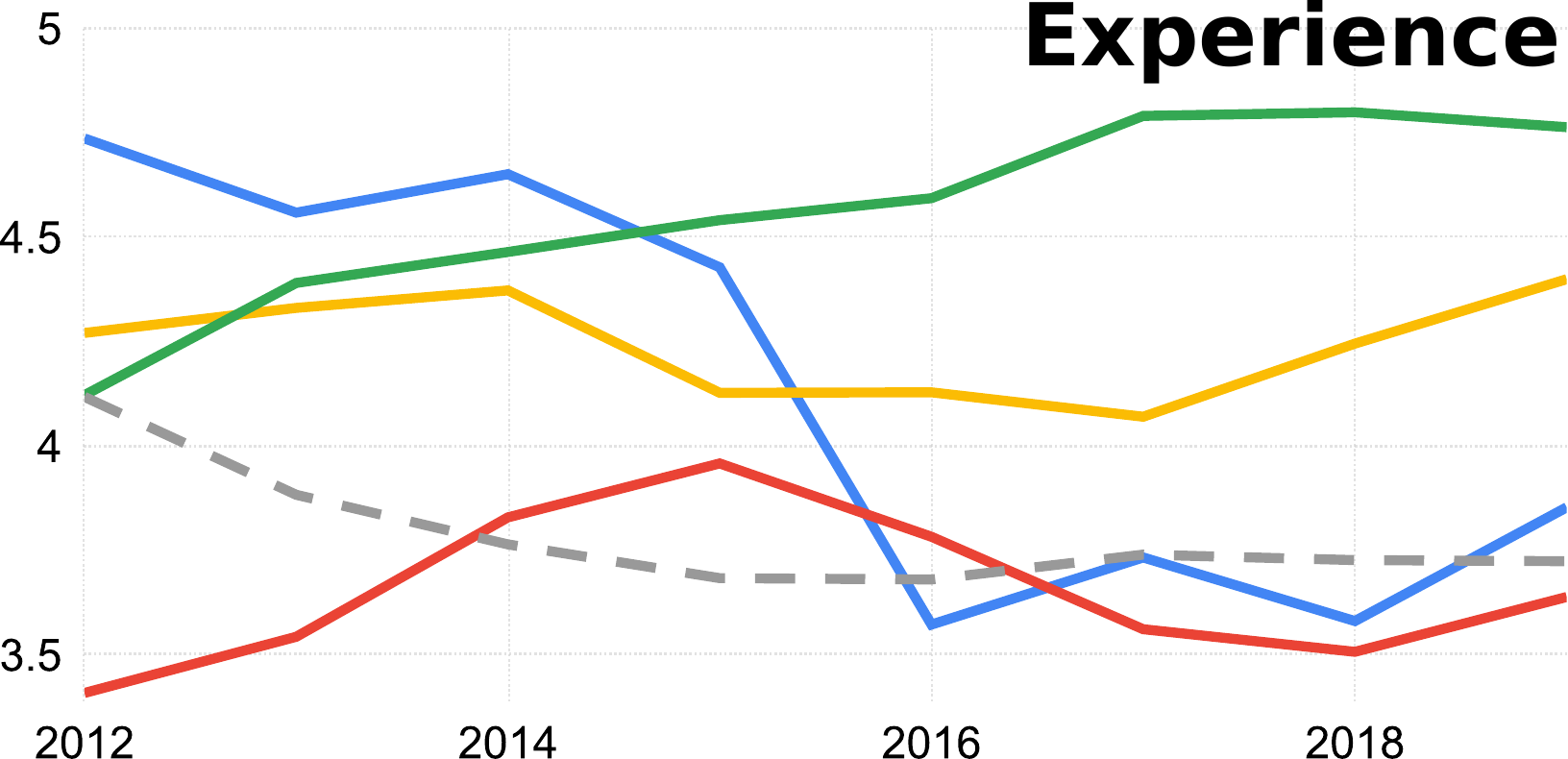}
		\label{subfig:experience}
	}
	\subfloat[]{
		\includegraphics[height = \myheightB\textheight]{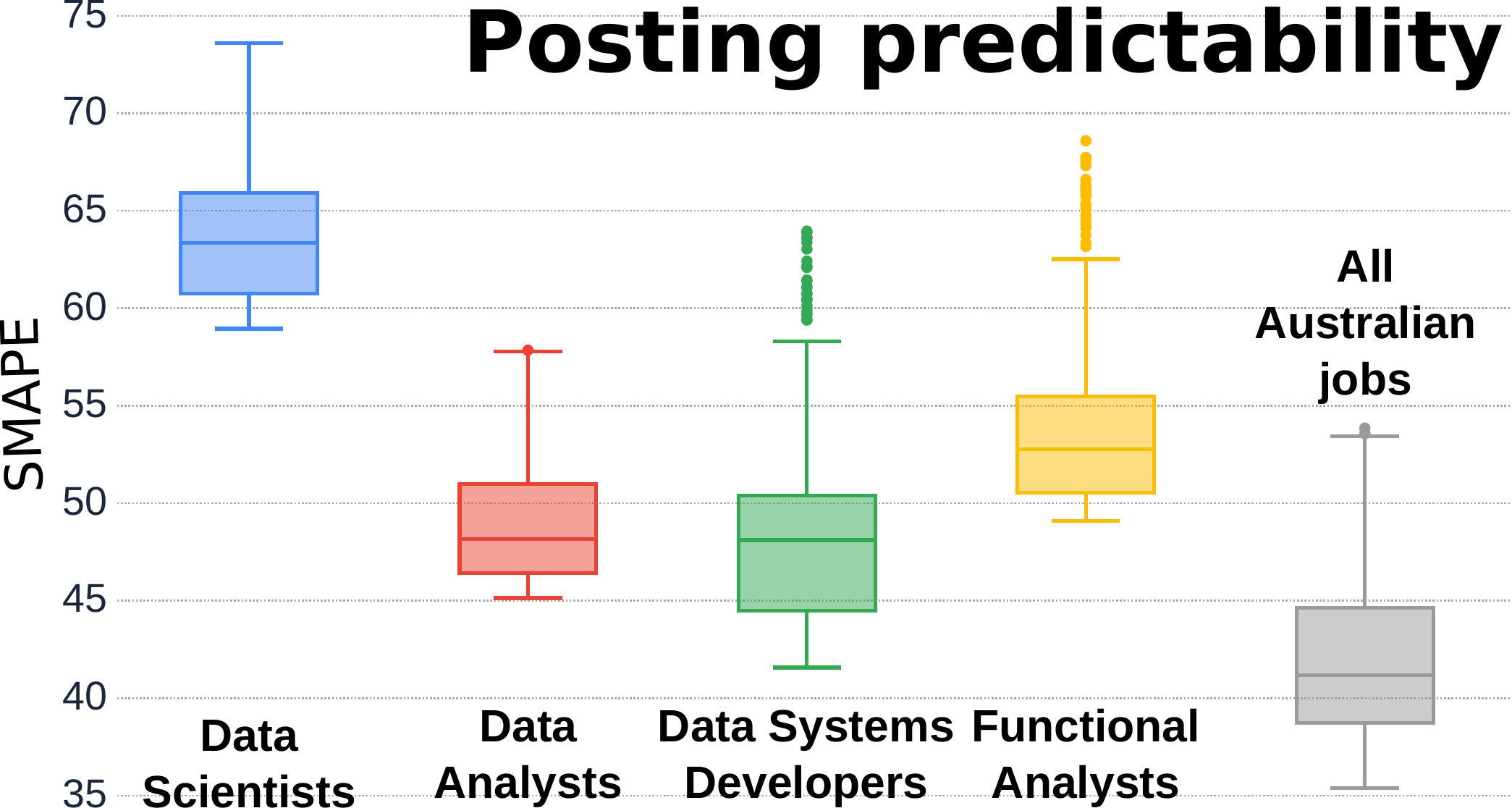}
		\label{subfig:post-predictability}
	}
	\caption{
		\textbf{Labour demand variables for detecting skill shortages from job ads data}:
		posting frequency \textbf{(a)} and its annual growth \textbf{(b)}; 
		median salary (Australian \$) \textbf{(c)}; 
		education level (years of formal education) \textbf{(d)}; 
		experience (years) \textbf{(e)} and 
		job ad posting predictability in terms of SMAPE error scores \textbf{(f)}.
	}
	\label{fig:variables}
\end{figure*}

\subsection{Predict job ad posting}
\label{subsec:job-ad-prediction}

\textbf{Forecast ad postings.}
In this section, we propose a `predictability' feature by building a time series model to predict job ad posting frequencies for each of the categories~\cite{appendix}. 
We use the Prophet time series forecasting tool developed by Facebook Research~\cite{taylor2018forecasting}. 
Prophet is an auto-regressive tool that fits non-linear time series trends with the effects from daily, weekly, and yearly seasonality, and also holidays. 
The three main model components are represented in the following equation:
\begin{equation} \label{eq:prophet}
    y(t) = g(t) + s(t) + h(t) + \epsilon_t    
\end{equation}
\noindent 
where $g(t)$ refers to the trend function that models non-periodic changes over time; $s(t)$ represents periodic changes, such as seasonality; $h(t)$ denotes holiday effects; and $\epsilon_t$ is the error term and represents all other idiosyncratic changes.
For more details on Prophet and its hyper-parameter choices, please refer to the online appendix~\cite{appendix}.

\textbf{Prediction error measure.}
Using \cref{eq:prophet}, one can run forward time and forecast job ad posting frequency.
We measure the accuracy of the forecast using the Symmetric Mean Absolute Percentage Error (SMAPE)~\cite{Scott_Armstrong1985-tt, makridakis1993accuracy}. SMAPE is formally defined as:
\begin{equation*}
    SMAPE(A_t,F_t) = \frac{200}{T} \sum_{t=1}^{T} \frac{|F_t - A_t|}{(|A_t|+|F_t|)}
\end{equation*}
\noindent
where $A_t$ denotes the actual value of jobs posted on day $t$, and $F_t$ is the predicted value of job ads on day $t$. SMAPE ranges from 0 to 200, with 0 indicating a perfect prediction and 200 the largest possible error. When actual and predicted values are both 0, we define SMAPE to be 0. We selected SMAPE as an alternative to MAPE because it is (1) scale-independent and (2) can handle actual or predicted zero values.
For a discussion on alternate error metrics, please consult the online appendix~\cite{appendix}.

\textbf{Evaluation protocol.}
The forecasts made using Prophet are deterministic (i.e. given the same input, we will obtains the same output).
We evaluate the uncertainty of predicted future job ad volumes using a `sliding window' approach. 
As shown in \cref{fig:sliding-window}, we use a constant number of training days (here $1,186$ days) to train the model, and we test the forecasting performance on the next $365$ days.
We shift both the training and the testing periods right by one day, and we repeat the process.
We iterate this process 365 times, denoted in \cref{fig:sliding-window} using \textit{Train start} for the starting point of the train period, \textit{Test start} for the starting point of the test period, and using \textit{Window start} for the starting point of the unused period.
Consequently, we train and test the model 365 times, and we obtain 365 SMAPE performance values, which are presented aggregated as a boxplot in \cref{subfig:post-predictability}.
The advantage of this approach is that it provides a distribution of SMAPE scores across a range of testing periods, which allows for a more robust evaluation of the modelling performance.

\begin{figure}[htp]
    \centering
    \includegraphics[width=.45\textwidth]{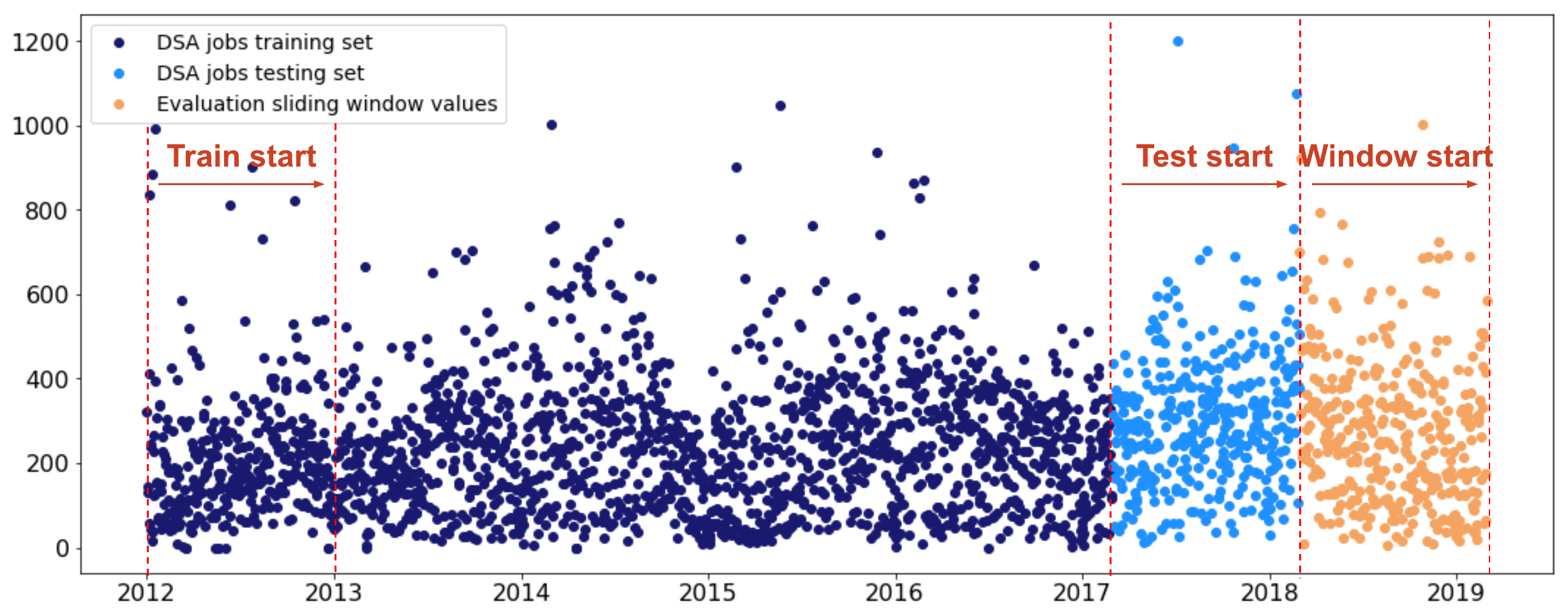}
    \caption{Sliding window setup for evaluating job ads forecasting performance.}
    \label{fig:sliding-window}
\end{figure}


\section{Discussion}

\begin{figure}[tbp]
    \centering
    \includegraphics[width=.45\textwidth]{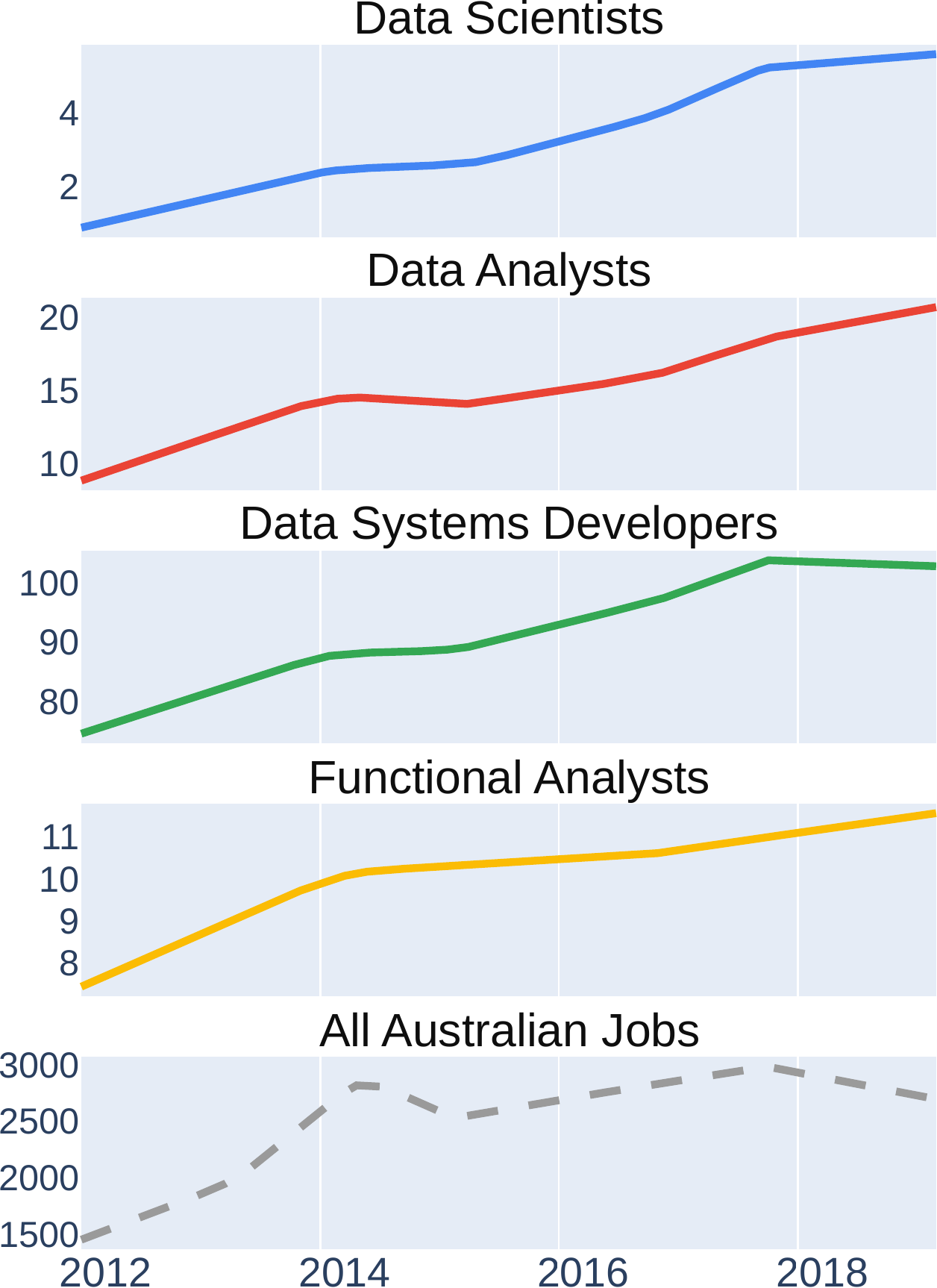}
    \caption{Trend lines of daily online job ad postings}
    \label{fig:trend-lines}
\end{figure}

Job ad posting trends ($g(t)$ in \cref{eq:prophet}) have grown for all DSA categories since 2012.
This is shown in \cref{fig:trend-lines}, which isolates $g(t)$ for each category to highlight the non-periodic changes of daily job ad posts.
Here, \cref{fig:trend-lines} shows that the more technically rigorous categories of `Data Scientists' and `Data Analysts' have experienced the highest growth trends. There are three distinct change point periods observed in \cref{fig:trend-lines}. Firstly, from January 2012 to April 2014, where the frequency of all job ads are growing.  Over this period, only `Data Scientists' grew at a faster rate than the total market for `All Australian Job' Ads (using the simple growth formula). This period can perhaps be explained by (1) the higher levels of job openings being posted online earlier in the dataset and (2) the early stages of DSA skills demanded by occupations, particularly for the more technically rigorous occupations.

The second period, from approximately May 2014 to November 2017, was generally one of slowing growth for online job ads. A possible explanation for this period is Australia's increasing underemployment rate~\cite{Australian_Bureau_of_Statistics2018-ck}. Underemployment rose relatively steeply from just above 7\% in 2014, diverging from a lowering unemployment rate, before reaching a peak just below 9\% around the beginning of 2017. Underemployment then began to slightly decrease until the end of 2018. The sharp rise in underemployment could be indicative of employers being less willing or able to hire due to softening labour market conditions, which would presumably affect the frequency of job ad postings. While the more analytically rigorous categories of `Data Scientists' and `Data Analysts' also experienced slowing growth, they both grew at higher rates relative to other categories. The fact that these categories maintained strong upward trends, despite dampening labour market forces, highlights the high levels of labour demand for these occupational categories.

The final period from October 2017 until February 2019 (the end of this dataset), was generally one of stagnation or slight growth. Again, `Data Scientists' and `Data Analysts' continued upward trajectories, albeit at slower growth rates than previous periods. All DSA categories had higher trend growth rates than `All Australian Job Postings' during this period. This final change point period highlights some possible conclusions. Firstly, the frequency of online job ads have potentially reached a saturation point. This means that the maximum proportion of job postings captured via online aggregators might have reached its upper limits. If this is the case, then any posting frequency growth for specific occupational classes above the total market rate could indicate high (or relatively high) labour demand. From this perspective, all DSA jobs continue to experience higher labour demands relative to all Australian job ads postings in the dataset since 2014.

The strong relative growth of `Data Scientists' and `Data Analysts' also provides insight. One interpretation is that Australian firms and employers have started to increasingly adopt AI technologies. A recent report by McKinsey \& Co suggests that this is the case~\cite{Taylor2019-al}. The accelerating rate of AI adoption requires highly skilled labour to make productive use of these technologies. These are the same analytically rigorous skills that are demanded from `Data Scientists' and `Data Analysts'. As a result, some portion of this growing labour demand for DSA skills, particularly the highly technical DSA skills, could be explained by accelerating AI adoption by Australian firms. Another related perspective is that Australian firms have increasing access to data with potentially meaningful insights. Therefore, workers with DSA skills that are able to productively use and draw insights from such data would logically be in high demand.

\section{Conclusions and Future Research}

In this research, we firstly developed a data-driven methodology to construct an adaptive set of skills highly similar to a set of seed skills. We then applied this method to identify the DSA skills set and DSA occupations, organising these occupations into common DSA categories. Secondly, we proposed five variables from online job ads data which are critical for the real-time detection of skill shortages. We then analysed the DSA categories according to each of these five variables. Here, we find strong evidence for how these features are collectively predictive of skill shortages. From this analysis, we find evidence that Australia is experiencing skills shortages for `Data Scientists' and `Data Analysts' occupations. A combination of indicators points to these conclusions. Firstly, both categories have experienced high relative growth in terms of \textbf{job ad posting frequencies}. Secondly, both categories command high, and growing, \textbf{wage premiums}. Thirdly, both categories demand higher than average \textbf{education requirements}, which constrains the supply of skilled labour pursuing these vocations. This is especially the case for `Data Scientists'. Fourthly, the average minimum years of \textbf{experience} required by employers for these categories are low. For `Data Scientists', the minimum experience requirements have decreased by almost one year since 2012 and sit just above the market average. For `Data Analysts', the average years of minimum experience have been below the market average since 2017. Lastly, these occupational categories are relatively \textbf{difficult to predict}, especially for occupations in the `Data Scientists' category. Taken collectively, these factors form a strong case that the Australian labour market has been experiencing a shortage of `Data Scientists' and `Data Analysts'.

\textbf{Limitations and future work.} 
A limitation of this work is that it only consists of labour demand data, and estimates labour supply via the proxy of the five proposed variables.
Future work will corroborate these findings according to official labour shortage lists published by governments (i.e. a labour supply `ground truth'). 
This could be achieved by developing a multivariate logistic classifier where the five proposed variables are used as features to predict whether an occupation is experiencing shortage. 
Conducting equivalent analyses on other markets and occupational groups could also provide insights into the predictive performance of these explanatory variables.

{\small
\section*{Acknowledgments}
Marian-Andrei Rizoiu was partially funded by the Science and Industry Endowment Fund, under project no. \textit{D61 Challenge: E06}.
Mary-Anne Williams was partially funded by the Australian Research Council Discovery under Discovery Project no. \textit{DP160102693}.
We would like to thank \textit{Burning Glass Technologies} for generously providing the data for this research.
}

\Urlmuskip=0mu plus 1mu\relax

\bibliographystyle{plain}
\bibliography{bibliography}

\input{supplemental} 

\end{document}

%% file: supplemental.tex
%
%
\clearpage
\appendix
\etocdepthtag.toc{mtappendix}
\etocsettagdepth{mtchapter}{none}
\etocsettagdepth{mtappendix}{subsection}
\etoctocstyle{1}{Contents (Appendix)}
\tableofcontents


This document is accompanying the submission \textit{\titlename}.
The information in this document complements the submission, and it is presented here for completeness reasons.
It is not required for understanding the main paper, nor for reproducing the results.

\subsection{Australia's looming DSA Shortfall}
The Australian Computer Society (ACS), Australia's peak body group for Information and Communication Technologies (ICTs), forecasts that Australia will need almost 100,000 additional ICT professionals just to keep up with demand by 2023~\cite{Deloitte_Access_Economics2018-dh}. Approximately half of these ICT professionals will require highly technical or digital management skills. However, domestic completions of ICT degrees were just 5,502 in 2016~\cite{Australian_Federal_Department_of_Education_and_Training2018-te}. This current level of labour supply is insufficient to meet the future demands for ICT professionals generally, and DSA occupations specifically.

\subsection{Limitations of Online Job Ads Data}
The biases discussed in \cref{related-work} are present in the dataset used for this research. For example, 52.8\% of Australian job ads in the dataset were classified as `Professionals' or `Managers' in 2018 (39.5\% and 13.3\%, respectively), according to the official Australian and New Zealand Standard Classification for Occupations (ANZSCO). These are typically `white collar' occupations. In comparison, employment data from the Australian Bureau of Statistics (ABS) indicates that `Professionals' and `Managers' collectively represent just 36.2\% of employment in Australia (23.7\% and 12.5\%, respectively)~\cite{Australian_Bureau_of_Statistics2019-sv}. The traditionally `blue collar' workers from categories such as `Machinery Operators and Drivers' and `Labourers' appear to be underrepresented in the BGT dataset.

Similarly, the 2018 average salary range for all online job ads in Australia was AUD\$89,028 - \$98,904. This is higher than the average full-time wage in Australia, which was \$83,408 in November 2018~\cite{Australian_Bureau_of_Statistics2018-zx}. Therefore, as online job ads fail to cover the universe of employment vacancies, they should be interpreted as trends rather than `ground truth’ for labour demand. However, these biases do not impede this research too significantly, as a major component of this research is comparing different classes of DSA jobs, which are all considered in the `Professionals' or `Managers' classes. 

\subsection{Challenges with Classifying Occupations}
A general challenge with classifying job ads is that job titles are not uniform. A `Senior Machine Learning Engineer' and a `Deep Learning Specialist' have different job titles but may require the same skills. Therefore, they should be measured in the same occupational class. An issue with ANZSCO, however, is that classifications are rarely updated; the last review was in 2013. So, emerging skills are not always properly captured or can be missed entirely leading to inaccurate classifications. So, the two example occupations above may be classified into different occupational classes despite having consistent skill requirements. Misclassified occupations can distort true representations of labour markets. Additionally, emerging skills, such as many DSA skills, complicate static and rarely updated occupational classifications.

These challenges have led BGT to develop their own taxonomies of labour skills and occupational classifications. BGT currently maintain a dictionary of over 11 thousand job skills. When processing job ads, BGT extract the skill requirements for each job. Typically there are multiple skill requirements for a unique job. For example, a `Data Science' job could consist of the following skills: `Python', `SQL', `Data Warehousing', `Communication Skills', and `Team Work / Collaboration'. These job skills build the foundation of BGT’s adaptive occupational classification system.

\subsection{DSA Skill Demands}
\begin{figure}[htp]
    \centering
    \includegraphics[width=8cm]{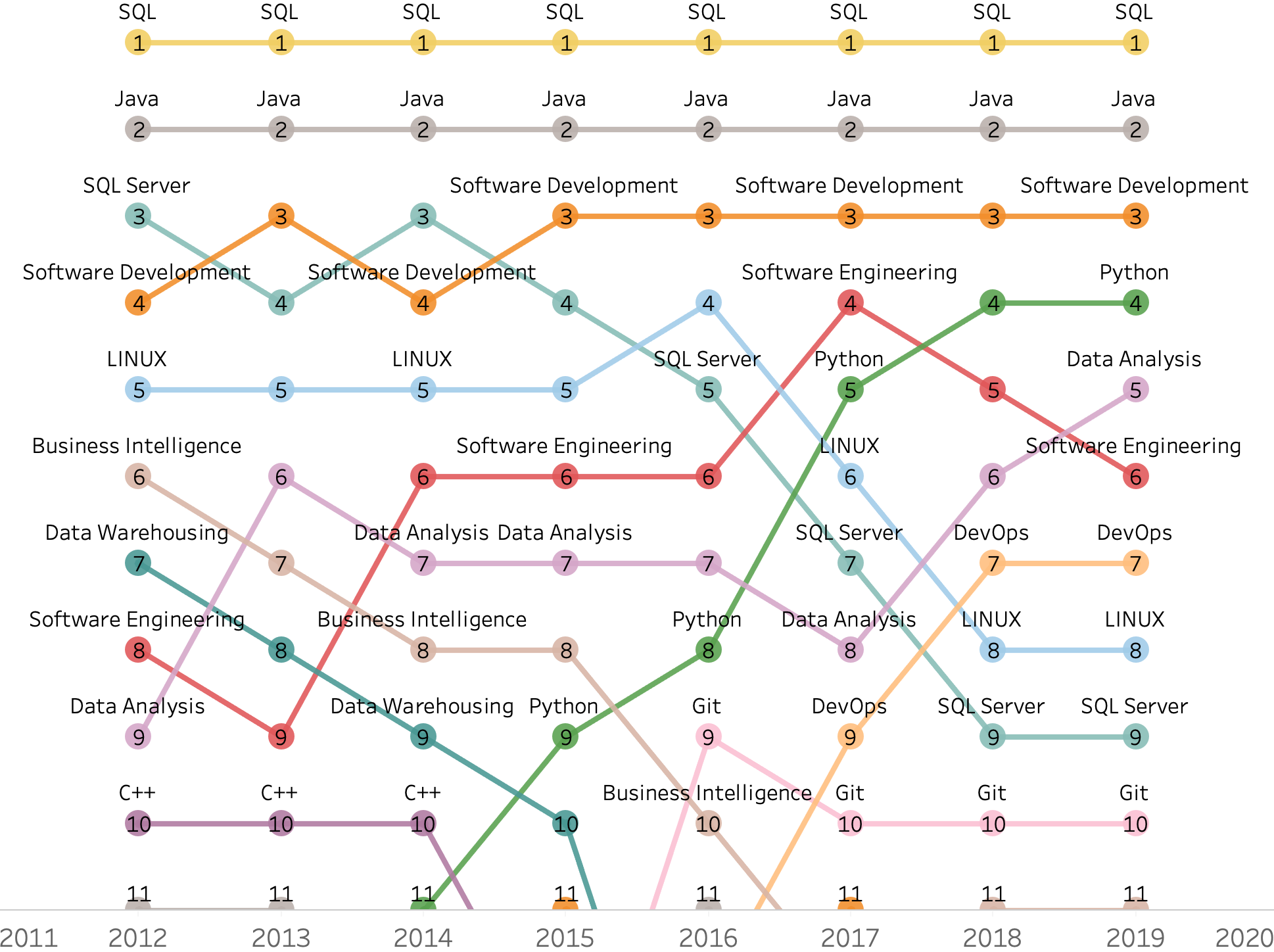}
    \caption{Top 10 DSA Skills for each year from 2012-2019}
    \label{fig:top-10-dsa}
\end{figure}

Comparing relative DSA skill demands involved counting the frequency of each DSA skill that occurs in unique DSA job ads. 
As seen in \cref{fig:top-10-dsa}, Structured Query Language (`SQL') has consistently been the DSA skill in the highest demand.

The Compound Annual Growth Rate (CAGR) was calculated for each DSA skill based on their first available posting in a DSA job, which is shown in \cref{tab:table2}.
`Blockchain' has unequivocally been the fastest growing DSA skill. However, this growth has been over a short period of time, with its first recording in 2016. 
The other fastest growing DSA skills have generally been analytical tools used either to manage `Big Data' or `Artificial Intelligence' (AI) related techniques. For instance, `Apache' Spark', `Apache Kafka', and [Apache] `PIG' are all open source software tools used to assist with `Big Data' management and processing. Additionally, skills such as `TensorFlow', `Deep Learning', and `Random Forests' are all skills that generally pertain to AI.

\begin{table}[h!]
	\centering
    \caption{Top DSA Skills Growth}
    \label{tab:table2}
    \begin{tabular}{cp{5.5cm}r}
      \toprule
      \multicolumn{1}{c}{\textbf{Rank}} & 
      \multicolumn{1}{c}{\textbf{DSA Skill}} &
      \multicolumn{1}{c}{\textbf{CAGR}}\\
      \midrule
      1	& Blockchain & 616\%\\
      2 & TensorFlow & 283\%\\
      3 & Apache Spark & 271\%\\
      4 & Deep Learning & 201\%\\
      5 & Apache Kafka & 188\%\\
      6 & Internet of Things (IoT) & 182\%\\
      7 & Microsoft Power BI & 175\%\\
      8 & Data Lakes / Reservoirs & 169\%\\
      9 & Qlik & 157\%\\
      10 & Random Forests & 151\%\\
      11 & Apache Hive & 145\%\\
      12 & PIG & 136\%\\
      13 & Pipeline (Computing) & 134\%\\
      14 & Supervised Learning (Machine Learning) & 131\%\\
      15 & Boosting (Machine Learning) & 129\%\\
      16 & Alteryx & 128\%\\
      17 & Sqoop & 119\%\\
      18 & Apache Flume & 109\%\\
      19 & DevOps & 107\%\\
      20 & Unsupervised Learning & 102\%\\
      \bottomrule
    \end{tabular}
\end{table}

\subsection{Time Series Forecasting with Prophet}
Time series analysis provides a set of techniques to draw inferences from a sequence of observations stored in time order~\cite{brockwell2002introduction}. The development of accurate time series models can offer insights into the principal components that have affected historical growth trajectory patterns. They also facilitate a means for making predictions into the future.

This paper applies a relatively new and high-performing time series forecasting tool developed by Facebook, called Prophet~\cite{taylor2018forecasting}. The forecasting tool is applied to Australian online job ads data to uncover growth trends of DSA jobs.

In 2017, Facebook Research released Prophet as an open source forecasting procedure implemented in the Python and R programming languages. When benchmarked against ARIMA, ETS (error, trend, seasonality) forecasting, seasonal naive forecasting, and the TBATS model, Prophet forecasts had significantly lower Mean Absolute Percentage Errors (MAPE)~\cite{taylor2018forecasting}. 



The default hyperparameters of Prophet were applied for this analysis. This included an uncertainty interval of 80\%, the automatic detection of trend change points, and the estimations of seasonality using a partial Fourier sum. For seasonality, Prophet uses a Fourier order of 3 for weekly seasonality and 10 for yearly seasonality. Experimentation steps were conducted by specifying a custom holidays dataframe, adjusting smoothing parameters, and fitting the model with a multiplicative seasonality setting. However, all of these specifications led to a slight deterioration of performance metrics. Therefore, the default hyperparameters were restored, which the authors state \textit{``are appropriate for most forecasting problems"}\cite{taylor2018forecasting}.

\subsection{Evaluating performance}
The Prophet library includes a method for calculating a range of evaluation metrics.\footnote{The method is called \texttt{cross\_validation}. For more information, see: https://facebook.github.io/prophet/docs/diagnostics.html} However, these metrics are not ideal for measuring prediction performance of online job ads for two reasons. 

Firstly, analyses in this paper are comparing DSA categories with different scales of job posting frequencies. Therefore, most metrics calculated by Prophet's diagnostics method, such as Mean Squared Error (MSE), Mean Absolute Error (MAE), and Root Mean Square Error (RMSE), are not suitable for comparisons because such measurements are scale-dependant~\cite{de200625}. 

Secondly, an appropriate performance metric for this dataset must not be distorted by zero values. This is important for job posts, where some DSA categories recorded zero daily postings, particularly earlier in the dataset. Subsequently, this rules out the last meaningful performance metric calculated by Prophet’s diagnostics, namely MAPE. As the dataset contains zero values for posting frequencies, MAPE values can be infinite as it involves division by zero.

Therefore, accommodating for these two criterion points, the selected prediction performance metric is the Symmetric Mean Absolute Percentage Error (SMAPE). SMAPE is an alternative to MAPE that is (1) scale-independent and (2) can handle actual or predicted zero values. SMAPE, first proposed by Armstrong~\cite{Scott_Armstrong1985-tt} and then by Makridakis~\cite{makridakis1993accuracy}, 

\subsection{DSA Skills List}
\cref{tab:DSA-skills} shows the selected DSA skills, i.e. the top 150 skills selected using the methodology described in \cref{sec:skills-similarity}.

\begin{table*}[p]
\caption{Selected 150 Data Science and Analytics skills.}
\setlength{\tabcolsep}{2pt}

\begin{tabular}{ll}

\begin{tabular}{lll}
\textbf{Rank} & \textbf{Skill}                                                   & \textbf{Theta}       \\
1    & Machine Learning                                        & 0.375157109 \\
2    & Data Science                                            & 0.339677644 \\
3    & Big Data                                                & 0.281395532 \\
4    & Data Mining                                             & 0.275784695 \\
5    & Artificial Intelligence                                 & 0.268911214 \\
6    & Apache Hadoop                                           & 0.160263705 \\
7    & R                                                       & 0.120578077 \\
8    & Big Data Analytics                                      & 0.11683186  \\
9    & Predictive Models                                       & 0.087256126 \\
10   & Scala                                                   & 0.078168962 \\
11   & Tableau                                                 & 0.071103958 \\
12   & Apache Hive                                             & 0.068540161 \\
13   & Python                                                  & 0.067852169 \\
14   & SAS                                                     & 0.058335431 \\
15   & NoSQL                                                   & 0.054171879 \\
16   & Teradata                                                & 0.053266061 \\
17   & SPSS                                                    & 0.052294251 \\
18   & Natural Language Processing                             & 0.051589073 \\
19   & MATLAB                                                  & 0.049969987 \\
20   & Data Visualisation                                      & 0.049141083 \\
21   & Data Transformation                                     & 0.043785348 \\
22   & MapReduce                                               & 0.04200936  \\
23   & Data Modelling                                          & 0.041207512 \\
24   & Statistical Analysis                                    & 0.040950811 \\
25   & Predictive Analytics                                    & 0.040725603 \\
26   & Statistics                                              & 0.040600659 \\
27   & Deep Learning                                           & 0.040097617 \\
28   & Internet of Things (IoT)                                & 0.038865379 \\
29   & PIG                                                     & 0.038346523 \\
30   & Extraction Transformation and Loading (ETL)             & 0.037375468 \\
31   & Data Architecture                                       & 0.037357392 \\
32   & Data Warehousing                                        & 0.037120923 \\
33   & Microsoft Power BI                                      & 0.03691897  \\
34   & Apache Kafka                                            & 0.03478849  \\
35   & Neural Networks                                         & 0.034594775 \\
36   & Data Engineering                                        & 0.033870742 \\
37   & Econometrics                                            & 0.033635451 \\
38   & Data Integration                                        & 0.031413571 \\
39   & Data Structures                                         & 0.029579863 \\
40   & Decision Trees                                          & 0.029538939 \\
41   & Business Intelligence                                   & 0.028968279 \\
42   & C++                                                     & 0.028931884 \\
43   & Pipeline (Computing)                                    & 0.027558689 \\
44   & Consumer Behaviour                                      & 0.0273288   \\
45   & Hadoop Cloudera                                         & 0.027221747 \\
46   & Data Quality                                            & 0.0264852   \\
47   & Clustering                                              & 0.026032976 \\
48   & Apache Webserver                                        & 0.026020174 \\
49   & Qlikview                                                & 0.025944556 \\
50   & Cassandra                                               & 0.025060662 \\
51   & Consumer Research                                       & 0.024973131 \\
52   & Apache Spark                                            & 0.024017603 \\
53   & AWS Redshift                                            & 0.023822744 \\
54   & Data Manipulation                                       & 0.023299597 \\
55   & Cluster Analysis                                        & 0.022795077 \\
56   & Microsoft Azure                                         & 0.022690165 \\
57   & Experiments                                             & 0.022525239 \\
58   & Physics                                                 & 0.021968001 \\
59   & Software Engineering                                    & 0.020672929 \\
60   & Cloud Computing                                         & 0.020237968 \\
61   & MongoDB                                                 & 0.020228716 \\
62   & Consumer Segmentation                                   & 0.0202243   \\
63   & DevOps                                                  & 0.020103595 \\
64   & Relational Databases                                    & 0.01974885  \\
65   & Data Analysis                                           & 0.019621418 \\
66   & Blockchain                                              & 0.019568638 \\
67   & Data Governance                                         & 0.019300535 \\
68   & SQL                                                     & 0.019192807 \\
69   & SQL Server Analysis Services (SSAS)                     & 0.018858212 \\
70   & Java                                                    & 0.018541708 \\
71   & TensorFlow                                              & 0.018237584 \\
72   & Text Mining                                             & 0.017501842 \\
73   & Random Forests                                          & 0.0173648   \\
74   & Robotics                                                & 0.01663332  \\
75   & Distributed Computing                                   & 0.01659359  \\
\end{tabular} &

\begin{tabular}{lll}
\textbf{Rank} & \textbf{Skill}                                                   & \textbf{Theta}       \\
76   & Computer Vision                                         & 0.016534028 \\
77   & Ruby                                                    & 0.016521212 \\
78   & Microsoft Sql Server Integration Services (SSIS)        & 0.016224833 \\
79   & PostgreSQL                                              & 0.015755516 \\
80   & Informatica                                             & 0.015750079 \\
81   & Applied Statistics                                      & 0.014990736 \\
82   & SQL Server Reporting Services (SSRS)                    & 0.01460998  \\
83   & Data Management                                         & 0.014488424 \\
84   & Data Lakes / Reservoirs                                 & 0.014444455 \\
85   & Metadata                                                & 0.014422194 \\
86   & Quantitative Analysis                                   & 0.014245931 \\
87   & Qlik                                                    & 0.013849961 \\
88   & ElasticSearch                                           & 0.013784912 \\
89   & Information Retrieval                                   & 0.013626625 \\
90   & Scalability Design                                      & 0.013495411 \\
91   & Database Design                                         & 0.013409781 \\
92   & Apache Flume                                            & 0.013268289 \\
93   & Supervised Learning (Machine Learning)                  & 0.013255296 \\
94   & Regression Algorithms                                   & 0.013068441 \\
95   & Model Building                                          & 0.012974866 \\
96   & Visual Basic for Applications (VBA)                     & 0.012941596 \\
97   & PERL Scripting Language                                 & 0.012885431 \\
98   & Cognos Impromptu                                        & 0.012817815 \\
99   & SAP BusinessObjects                                     & 0.012601388 \\
100  & Oracle Business Intelligence Enterprise Edition (OBIEE) & 0.012256767 \\
101  & Prototyping                                             & 0.012183407 \\
102  & Node.js                                                 & 0.012089477 \\
103  & Experimental Design                                     & 0.012083924 \\
104  & MySQL                                                   & 0.012051979 \\
105  & Classification Algorithms                               & 0.01192503  \\
106  & Logistic Regression                                     & 0.011923395 \\
107  & Relational DataBase Management System (RDBMS)           & 0.011907611 \\
108  & Statistical Methods                                     & 0.011798527 \\
109  & Splunk                                                  & 0.0116979   \\
110  & Sqoop                                                   & 0.011619513 \\
111  & GitHub                                                  & 0.011606854 \\
112  & Unsupervised Learning                                   & 0.011432418 \\
113  & Apache Impala                                           & 0.011420459 \\
114  & Web Analytics                                           & 0.011406332 \\
115  & Git                                                     & 0.011202096 \\
116  & Amazon Web Services (AWS)                               & 0.01118572  \\
117  & Datastage                                               & 0.011123658 \\
118  & Optimisation                                            & 0.011085172 \\
119  & Simulation                                              & 0.010785033 \\
120  & LINUX                                                   & 0.010773868 \\
121  & Software Development                                    & 0.010750719 \\
122  & Continuous Integration (CI)                             & 0.010688564 \\
123  & Business Intelligence Reporting                         & 0.010349562 \\
124  & Agile Development                                       & 0.010225424 \\
125  & Solution Architecture                                   & 0.010225063 \\
126  & AWS Elastic Compute Cloud (EC2)                         & 0.010217691 \\
127  & Microstrategy                                           & 0.010147521 \\
128  & Marketing Analytics                                     & 0.010006654 \\
129  & Bash                                                    & 0.009937595 \\
130  & Alteryx                                                 & 0.009881429 \\
131  & SQL Server                                              & 0.009830543 \\
132  & Shell Scripting                                         & 0.009614866 \\
133  & Credit Risk                                             & 0.009534963 \\
134  & Image Processing                                        & 0.009483378 \\
135  & Boosting (Machine Learning)                             & 0.009409621 \\
136  & Platform as a Service (PaaS)                            & 0.009390802 \\
137  & Transact-SQL                                            & 0.009342661 \\
138  & Version Control                                         & 0.009182692 \\
139  & Support Vector Machines (SVM)                           & 0.009167358 \\
140  & Data Warehouse Processing                               & 0.00903522  \\
141  & Customer Acquisition                                    & 0.009029462 \\
142  & Linear Regression                                       & 0.008983594 \\
143  & Software Architecture                                   & 0.008952848 \\
144  & Google Analytics                                        & 0.008950648 \\
145  & AWS Simple Storage Service (S3)                         & 0.008939552 \\
146  & Dimensional and Relational Modelling                    & 0.008727614 \\
147  & Microsoft SQL                                           & 0.008714559 \\
148  & Functional Programming                                  & 0.008700033 \\
149  & Scrum                                                   & 0.008677026 \\
150  & Economics                                               & 0.008593447
\end{tabular}

\end{tabular}
\label{tab:DSA-skills}
\end{table*}